\documentclass[twoside,twocolumn,9pt]{article}
\usepackage{extsizes}
\usepackage[super,sort&compress,comma]{natbib} 
\usepackage[version=3]{mhchem}
\usepackage[left=1.5cm, right=1.5cm, top=1.785cm, bottom=2.0cm]{geometry}
\usepackage{balance}
\usepackage{times,mathptmx}
\usepackage{sectsty}
\usepackage{graphicx} 
\usepackage{lastpage}
\usepackage[format=plain,justification=justified,singlelinecheck=false,font={stretch=1.125,small,sf},labelfont=bf,labelsep=space]{caption}
\usepackage{float}
\usepackage{fancyhdr}
\usepackage{fnpos}
\usepackage[english]{babel}
\addto{\captionsenglish}{%
  
}
\usepackage{array}
\usepackage{droidsans}
\usepackage{charter}
\usepackage[T1]{fontenc}
\usepackage[usenames,dvipsnames]{xcolor}
\usepackage{setspace}
\usepackage[compact]{titlesec}
\usepackage{hyperref}

\usepackage{epstopdf}

\definecolor{cream}{RGB}{222,217,201}

\newcommand{\p}[1]{\left(#1\right)}

\usepackage{amsmath,amssymb}
\usepackage{mathrsfs}
\usepackage{mathtools}

\usepackage[disable]{todonotes} 				
\newcommand*{\blauw}[1]{{#1}}

\begin{document}

\pagestyle{fancy}
\thispagestyle{plain}
\fancypagestyle{plain}{

\renewcommand{\headrulewidth}{0pt}
}

\makeFNbottom
\makeatletter
\renewcommand\LARGE{\@setfontsize\LARGE{15pt}{17}}
\renewcommand\Large{\@setfontsize\Large{12pt}{14}}
\renewcommand\large{\@setfontsize\large{10pt}{12}}
\renewcommand\footnotesize{\@setfontsize\footnotesize{7pt}{10}}
\makeatother

\renewcommand{\thefootnote}{\fnsymbol{footnote}}
\renewcommand\footnoterule{\vspace*{1pt}%
\color{cream}\hrule width 3.5in height 0.4pt \color{black}\vspace*{5pt}} 
\setcounter{secnumdepth}{5}

\makeatletter 
\renewcommand\@biblabel[1]{#1}            
\renewcommand\@makefntext[1]%
{\noindent\makebox[0pt][r]{\@thefnmark\,}#1}
\makeatother 
\renewcommand{\figurename}{\small{Fig.}~}
\sectionfont{\sffamily\Large}
\subsectionfont{\normalsize}
\subsubsectionfont{\bf}
\setstretch{1.125} 
\setlength{\skip\footins}{0.8cm}
\setlength{\footnotesep}{0.25cm}
\setlength{\jot}{10pt}
\titlespacing*{\section}{0pt}{4pt}{4pt}
\titlespacing*{\subsection}{0pt}{15pt}{1pt}

\fancyfoot{}
\fancyfoot[LO,RE]{\vspace{-7.1pt}\includegraphics[height=9pt]{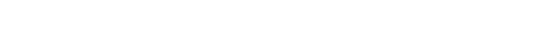}}
\fancyfoot[CO]{\vspace{-7.1pt}\hspace{13.2cm}\includegraphics{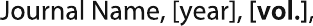}}
\fancyfoot[CE]{\vspace{-7.2pt}\hspace{-14.2cm}\includegraphics{RF}}
\fancyfoot[RO]{\footnotesize{\sffamily{1--\pageref{LastPage} ~\textbar  \hspace{2pt}\thepage}}}
\fancyfoot[LE]{\footnotesize{\sffamily{\thepage~\textbar\hspace{3.45cm} 1--\pageref{LastPage}}}}
\fancyhead{}
\renewcommand{\headrulewidth}{0pt} 
\renewcommand{\footrulewidth}{0pt}
\setlength{\arrayrulewidth}{1pt}
\setlength{\columnsep}{6.5mm}
\setlength\bibsep{1pt}

\makeatletter 
\newlength{\figrulesep} 
\setlength{\figrulesep}{0.5\textfloatsep} 

\newcommand{\topfigrule}{\vspace*{-1pt}%
\noindent{\color{cream}\rule[-\figrulesep]{\columnwidth}{1.5pt}} }

\newcommand{\botfigrule}{\vspace*{-2pt}%
\noindent{\color{cream}\rule[\figrulesep]{\columnwidth}{1.5pt}} }

\newcommand{\dblfigrule}{\vspace*{-1pt}%
\noindent{\color{cream}\rule[-\figrulesep]{\textwidth}{1.5pt}} }

\makeatother

\twocolumn[
  \begin{@twocolumnfalse}
\vspace{3cm}
\sffamily
\begin{tabular}{m{4.5cm} p{13.5cm} }

\includegraphics{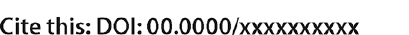} & \noindent\LARGE{\textbf{Molecular Latent Space Simulators}} \\
\vspace{0.3cm} & \vspace{0.3cm} \\

 & \noindent\large{Hythem Sidky,\textit{$^{a}$} Wei Chen,\textit{$^{b}$} and Andrew L. Ferguson\textit{$^{\ast}$$^{a}$}} \\

\includegraphics{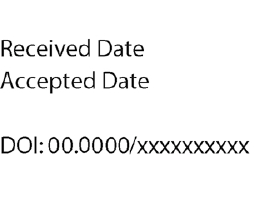} & \noindent\normalsize{Small integration time steps limit molecular dynamics (MD) simulations to millisecond time scales. Markov state models (MSMs) and equation-free approaches learn low-dimensional kinetic models from MD simulation data by performing configurational or dynamical coarse-graining of the state space. The learned kinetic models enable the efficient generation of dynamical trajectories over vastly longer time scales than are accessible by MD, but the discretization of configurational space and/or absence of a means to reconstruct molecular configurations precludes the generation of continuous all-atom molecular trajectories. We propose latent space simulators (LSS) to learn kinetic models for continuous all-atom simulation trajectories by training three deep learning networks to (i) learn the slow collective variables of the molecular system, (ii) propagate the system dynamics within this slow latent space, and (iii) generatively reconstruct molecular configurations. We demonstrate the approach in an application to Trp-cage miniprotein to produce novel ultra-long synthetic folding trajectories that accurately reproduce all-atom molecular structure, thermodynamics, and kinetics at six orders of magnitude lower cost than MD. The dramatically lower cost of trajectory generation enables greatly improved sampling and greatly reduced statistical uncertainties in estimated thermodynamic averages and kinetic rates.} \\

\end{tabular}

 \end{@twocolumnfalse} \vspace{0.6cm}

  ]

\renewcommand*\rmdefault{bch}\normalfont\upshape
\rmfamily
\section*{}
\vspace{-1cm}


\footnotetext{\textit{$^{a}$~Pritzker School of Molecular Engineering, University of Chicago, Chicago, USA. E-mail: andrewferguson@uchicago.edu}}
\footnotetext{\textit{$^{b}$~Department of Physics, University of Illinois at Urbana-Champaign, Urbana, USA. }}





\section{Introduction}

Molecular dynamics (\blauw{MD}) simulates the dynamical evolution of molecular systems by numerically integrating the classical equations of motion \cite{frenkel2002understanding}. Modern computer hardware \cite{stone2010gpu,Shaw2014} and efficient and scalable simulation algorithms \cite{phillips2005scalable,chow2008desmond,glaser2015strong,plimpton1993fast} have enabled the simulation of billion \cite{abraham2002work,abraham2002brittle} and trillion-atom systems \cite{tchipev2019twetris}. Advancing the barrier in time scale has proven far more challenging. Stability of the numerical integration requires time steps on the order of femtoseconds commensurate with the fastest atomic motions \cite{elber2016perspective}, which limits simulations to microseconds on commodity processors \cite{elber2016perspective} and milliseconds on special purpose hardware \cite{Shaw2014}. Enhanced sampling techniques apply accelerating biases and analytical corrections to recover thermodynamic averages \cite{torrie1977nonphysical,mcdonald1967machine,abrams2014enhanced,miao2016unconstrained,sidky2020machine} but -- except in special cases and the limit of small bias \cite{chodera2011dynamical,donati2018girsanov} -- no analogous approaches exist to recover unbiased dynamical trajectories from biased simulations.

The MD algorithm propagates a molecular configuration $\mathbf{x}_t$ at time $t$ to $\mathbf{x}_{t+\tau}$ via transition densities $\mathbf{x}_{t+\tau} \sim p_\tau(\mathbf{x}_{t+\tau} | \mathbf{x}_t)$ \cite{noe2018machine,Fernandez2020}. Assuming ergodicity, the probability density over microstates converges to the stationary distribution as $\displaystyle{\lim_{t \to \infty}} q_t(\mathbf{x}) = \pi(\mathbf{x})$. Breaking the time scale barrier requires a surrogate model for $p_\tau(\mathbf{x}_{t+\tau} | \mathbf{x}_t)$ that can be more efficiently evaluated and with larger time steps than MD. Accurately approximating this propagator in the high-dimensional $N$-atom configurational space $\mathbf{x} \in \mathbb{R}^{3N}$ is intractable. In general, for sufficiently large $\tau$ there is an emergent low-dimensional simplicity that admits accurate modeling of the dynamics by a low-dimensional propagator $p_\tau(\boldsymbol{\psi}_{t+\tau} | \boldsymbol{\psi}_t)$ within a latent space $\boldsymbol{\psi} \in \mathbb{R}^{m \ll 3N}$. The relation between MD and latent space dynamics can be represented as \cite{noe2018machine,Fernandez2020},
\begin{alignat}{4} 
&\mathbf{x}_t && \underset{E}{\rightarrow} \;\; && \boldsymbol{\psi}_t \notag \\
\mathrm{MD} &\downarrow{} &&   && \downarrow {\scriptstyle P} \label{eqn:scheme} \\
&\mathbf{x}_{t+\tau} && \underset{D}{\leftarrow} \;\; && \boldsymbol{\psi}_{t+\tau} \notag
\end{alignat} 
This scheme defines three learning problems \cite{noe2018machine}: (i) encoding $E$ of molecular configurations $\mathbf{x}$ to the latent space $\boldsymbol{\psi}$, (ii) propagation $P$ of the latent space dynamics according to transition densities $p_\tau(\boldsymbol{\psi}_{t+\tau} | \boldsymbol{\psi}_t)$, and (iii) decoding (or generating) $D$ of molecular configurations from the latent space \cite{noe2018machine}.

Markov state models (\blauw{MSM}) \cite{husic2018markov,pande2010everything,Prinz2011,bowman2013introduction,Sidky2019,Wehmeyer2019,Mardt2018,wu2017variational} and the equation-free approach of of Kevrekidis and co-workers \cite{kevrekidis2003equation,kevrekidis2004equation,kevrekidis2009equation,Mori1965,Zwanzig1973,zwanzig2001nonequilibrium,risken2012fokker} respectively employ configurational and dynamical coarse graining to parameterize low-dimensional propagators, but both methods lack molecular decoders. 
Recently, numerous deep learning approaches have been proposed to learn $E$, $P$, and $D$ from MD trajectories, including time-lagged autoencoders \cite{Wehmeyer2018}, time-lagged variational autoencoders \cite{Hernandez2017}, and time-lagged autoencoders with propagators \cite{Lusch2018}. Training these networks requires a time-lagged reconstruction term $|| \mathbf{x}_{t+\tau} - D \circ P \circ E(\mathbf{x}_t)||$ within the loss, which can cause the network to fail to approximate the true slow modes \cite{Chen2019}. Further, time-lagged autoencoders and time-lagged variational autoencoders do not learn valid propagators  \cite{noe2018machine}, and the inherent stochasticity of MD appears to frustrate learning of the propagator and decoder in time-lagged autoencoders with propagators \cite{noe2018machine}. Deep generative MSMs (\blauw{DeepGenMSM)} simultaneously learn a fuzzy encoding to metastable states and ``landing probabilities'' to decode molecular configurations \cite{Wu2018}. The method computes a proper propagator and generatively decodes novel molecular structures, but -- as with all MSM-based approaches -- it configurationally discretizes the latent space and relies on the definition of long-lived metastable states.

In this work, we propose molecular latent space simulators (\blauw{LSS}) as a means to train kinetic models over limited MD simulation data that are capable of producing novel all-atom molecular trajectories at orders of magnitude lower cost. The LSS can be conceived as means to augment conventional MD by distilling a kinetic model from training data, efficiently generating continuous all-atom trajectories, and computing high-accuracy estimates of any all-atom structural, thermodynamic, or kinetic observable. 

The LSS is based on three deep learning networks independently trained to (i) learn an encoding $E$ into a latent space of slow variables using state-free reversible VAMPnets (\blauw{SRV}) \cite{Chen2019a}, (ii) learn a propagator $P$ to evolve the system dynamics within this latent space using mixture density networks (\blauw{MDN}) \cite{Bishop1994,bishop2006pattern}, and (iii) learn a decoding $D$ from the latent space to molecular configurations using a conditional Wasserstein generative adversarial network (\blauw{cWGAN}) \cite{Gulrajani2017}. Separation of the learning problems in this manner makes training and deployment of the LSS modular and simple. The stochastic nature of the MDN propagator means that the trained kinetic model generates novel trajectories and does not simply recapitulate copies of the training data. The approach is distinguished from MSM-based approaches in that it requires no discretization into metastable states \cite{Fernandez2020,Wu2018}. The continuous formulation of the propagator in the slow latent space shares commonalities with the equation-free approach \cite{kevrekidis2003equation,kevrekidis2009equation}, but we eschew parameterizing a stochastic differential equation in favor of a simple and efficient deep learning approach, and also equip our simulator with a generative molecular decoder.


\section{Methods}

\begin{figure*}[ht!]
\begin{centering}
    \includegraphics[width=0.85\textwidth]{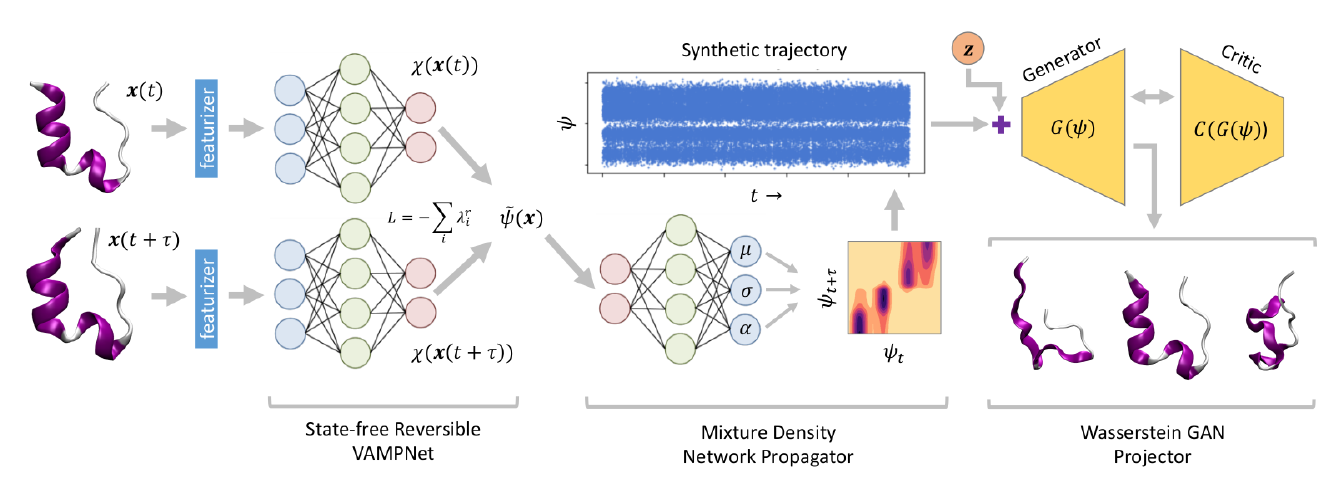}
    \caption{
    Schematic diagram of the latent space simulator (LSS) comprising three back-to-back deep neural networks. A state-free reversible VAMPnet (SRV) learns an encoding $E$ of molecular configurations into a latent space spanned by the leading eigenfunctions of the transfer operator (Eqn.~\ref{eqn:scheme}). A mixture density network (MDN) learns a propagator $P$ to sample transition probabilities $p_\tau(\boldsymbol{\psi}_{t+\tau} | \boldsymbol{\psi}_t)$ within the latent space. A conditional Wasserstein GAN (cWGAN) learns a generative decoding $D$ of molecular configurations conditioned on the latent space coordinates. The trained LSS is used to to generate ultra-long synthetic trajectories by projecting the initial configuration into the latent space using the SRV, sampling from the MDN to generate long latent space trajectories, and decoding to molecular configurations using the cWGAN.
	}
    \label{lss_diagram}
\end{centering}
\end{figure*}

A schematic diagram of the LSS and the three deep networks of which it is comprised is presented in Fig.~\ref{lss_diagram}. We describe each of the three component networks in turn and then describe LSS training and deployment.

\subsection{Encoder: State-free Reversible VAMPnets}

The transfer operator $\mathscr{T}$ at a lag time $\tau$ is the propagator of probability distributions over microstates with respect to the equilibrium density $u(\mathbf{x}) = q(\mathbf{x})/\pi(\mathbf{x})$ under transition densities $p_\tau(\mathbf{x}_{t+\tau} | \mathbf{x}_t)$ \cite{koltai2018optimal,klus2018data}. For sufficiently large $\tau$ the dynamics may be approximated as Markovian so $p_\tau(\mathbf{x}_{t+\tau} | \mathbf{x}_t)$ is time homogenous,
\begin{align}
u_{t+\tau}(\mathbf{x}) &= \mathscr{T} \circ u_t(\mathbf{x})=\frac{1}{\pi(\mathbf{x})} \int d\mathbf{y}\ p_\tau(\mathbf{x} | \mathbf{y})u_t(\mathbf{y})\pi(\mathbf{y}). \label{eqn:transfer}
\end{align}
In equilibrium systems obeying detailed balance $\pi(\mathbf{x})p_\tau(\mathbf{y} | \mathbf{x})=\pi(\mathbf{y})p_\tau(\mathbf{x} | \mathbf{y})$, $\mathscr{T}$ is identical to the Koopman operator, self-adjoint with respect to $\langle a\vert b\rangle_\pi=\int a(\mathbf{x})b(\mathbf{x})\pi(\mathbf{x})d\mathbf{x}$, and possesses a complete orthonormal set of eigenfunctions $\{\psi_i(\mathbf{x})\}$ with real eigenvalues $1 = \lambda_0 > \lambda_1 \geq \lambda_2 \geq \ldots$ \cite{Noe2013,Nuske2014,klus2018data,Chen2019a,wu2020variational},
\begin{align}
\mathscr{T} \circ \psi_i(\mathbf{x}) &= \lambda_i \psi_i(\mathbf{x}), \qquad \langle \psi_i\vert \psi_j\rangle_\pi=\delta_{ij}.
\end{align}
The pair ($\psi_0(\mathbf{x})$=$\mathbf{1}$,$\lambda_0$=$1$) corresponds to the equilibrium distribution at $t \rightarrow \infty$ and the remainder to a hierarchy of increasingly quicker relaxing processes with implied time scales $t_i = -\tau / \ln \lambda_i$ \cite{Chen2019a}. The evolution of $u_t(\mathbf{x})$ after $k$ applications of $\mathscr{T}$ is expressed in this basis as,
\begin{align}
u_{t+k\tau}(\mathbf{x}) = \mathscr{T}^k \circ u_t(\mathbf{x}) &= \sum_i  \langle \psi_i\vert u_t\rangle_\pi \exp\p{-\frac{k\tau}{t_i}} \psi_i(\mathbf{x}), \label{eqn:expansion}
\end{align}

The variational approach to conformational dynamics (\blauw{VAC}) defines a variational principle to approximate these eigenfunctions as $\tilde{\psi}_i(\mathbf{x}) = \sum_j s_{ij} \chi_j(\mathbf{x})$ within a basis $\{\chi_j\}$ by solving for optimal expansion coefficients $s_{ij}$ \cite{Noe2013,Nuske2014,Chen2019a}. SRVs \cite{Chen2019a} -- themselves based on VAMPnets, a deep learning-based method for MSM construction \cite{Mardt2018}, and closely related to extended dynamic mode decomposition with dictionary learning \cite{li2017extended} -- employ deep canonical correlation analysis (\blauw{DCCA}) \cite{Andrew2013} to learn both the optimal expansion coefficients \textit{and} optimal basis functions as nonlinear transformations of the (featurized) molecular coordinates. This is achieved by training twin-lobed deep neural networks to minimize a VAMP-r loss function $\mathscr{L}_\mathrm{SRV} = -\sum_{m}{\lambda_m^r}$ \cite{Mardt2018}. SRVs trained over MD trajectories furnish an encoding $E$ (Eqn.~\ref{eqn:scheme}) into a $m$-dimensional latent space spanned by $\{\psi_i(\mathbf{x})\}_{i=1}^m$, where $m$ is determined by a gap in the eigenvalue spectrum. This spectral encoding into the leading modes of $\mathscr{T}$ neglects fast processes with implied timescales $t_i << \tau$ (Eqn.~\ref{eqn:expansion}) and is an optimal parameterization of the system for a low-dimensional long-time propagator \cite{noe2018machine}.

\subsection{Propagator: Mixture Density Networks}

At sufficiently large $\tau$ the latent space $\boldsymbol{\psi}(\mathbf{x})$=$\{\psi_i(\mathbf{x})\}_{i=1}^m$ supports an autonomous dynamical system in the leading modes of $\mathscr{T}$. We train MDNs to learn transition densities $p_\tau(\boldsymbol{\psi}_{t+\tau} | \boldsymbol{\psi}_t)$ from MD trajectories projected in the latent space. MDNs combine deep neural networks with mixture density models to overcome poor performance of standard networks in learning multimodal distributions \cite{Bishop1994,bishop2006pattern}. Transition densities are approximated as a linear combination of $C$ kernels,
\begin{align}
p_\tau(\boldsymbol{\psi}_{t+\tau} | \boldsymbol{\psi}_t) &= \sum_{c=1}^{C} \alpha_c(\boldsymbol{\psi}_t) \boldsymbol{\phi}_c(\boldsymbol{\psi}_{t+\tau}; \boldsymbol{\mu}_c(\boldsymbol{\psi}_t), \boldsymbol{\sigma}_c(\boldsymbol{\psi}_t)), \label{eqn:mdn}
\end{align}
where we choose $\boldsymbol{\phi}_c$ to be $m$-dimensional Gaussians. The $\boldsymbol{\psi}_t$-dependent Gaussian means $\boldsymbol{\mu}_c(\boldsymbol{\psi}_t)$, variances $\boldsymbol{\sigma}_c(\boldsymbol{\psi}_t)$, and linear mixing coefficients $\alpha_c(\boldsymbol{\psi}_t)$ are learned by a deep feedforward neural network that minimizes the loss function $\mathscr{L}_\mathrm{MDN} = - \sum_{\gamma}{\ln p_\tau(\boldsymbol{\psi}_{t+\tau}^\gamma | \boldsymbol{\psi}_t^\gamma)}$, where $\gamma$ indexes pairs of time-lagged training data observations. The normalization $\sum_{c=1}^{C}{\alpha_c(\boldsymbol{\psi}_t)}$=1 is enforced by softmax activations and the $\boldsymbol{\mu}_c(\boldsymbol{\psi}_t)$ bounded using sigmoid activations.

The trained MDN defines the latent space propagator $P$ (Eqn.~\ref{eqn:scheme}) and we sample transition densities $p_\tau(\boldsymbol{\psi}_{t+\tau} | \boldsymbol{\psi}_t)$ to advance the system in time (Fig.~\ref{lss_diagram}). Propagation is conducted entirely within the latent space and does not require recurrent decoding and encoding to the molecular representations that can lead to accumulation of errors and numerical instability \cite{noe2018machine,Pathak2018}. The transition densities are learned from the statistics transitions in the training data and new trajectories are generated by sampling from these transition densities. These new trajectories therefore represent novel dynamical pathways over the latent space and are not simply recapitulations or approximate copies of those in the training data. Successful MDN training is contingent on the low-dimensional and Markovian nature of the latent space dynamics at large $\tau$ discovered by the SRVs.

\subsection{Decoder: Conditional Wasserstein GAN}

Generative adversarial networks are a leading neural network architecture for generative modeling \cite{Goodfellow2014}. We employ a cWGAN \cite{Arjovsky2017,Gulrajani2017} to decode from the latent space $\boldsymbol{\psi}$ to molecular configurations $\mathbf{x}$ by performing adversarial training between a generator $G(\mathbf{z})$ that outputs molecular configurations from inputs $\mathbf{z} \sim \mathcal{P}_z(\mathbf{z})$ and a critic $C(\mathbf{x})$ that evaluates the quality of a molecular configuration $\mathbf{x}$. The networks are jointly trained to minimize a loss function based on the Wasserstein (Earth Mover's) distance,
\begin{align}
\mathscr{L}_\mathrm{WGAN} &= \max_{w \in W} \mathbb{E}_{\mathbf{x} \sim \mathcal{P}_{x}}[C_w(\mathbf{x})] - \mathbb{E}_{\mathbf{z} \sim \mathcal{P}_{z}}[C_w(G(\mathbf{z}))],
\end{align}
where $\mathcal{P}_x(\mathbf{x})$ is the distribution over molecular configurations observed in the MD training trajectory and $\{C_w\}_{w \in W}$ is a family of $K$-Lipschitz functions enforced through a gradient penalty \cite{Arjovsky2017,Gulrajani2017}. To generate molecular configurations consistent with particular states in the latent space we pass $\boldsymbol{\psi}$ as a conditioning variable to $G$ and $C$ \cite{Mirza2014} and drive the generator with $d$-dimensional Gaussian noise $\mathcal{P}_z(\mathbf{z}) = \mathcal{N}(\mathbf{0},\mathbf{1}) \in \mathbb{R}^d$. The noise enables $G$ to generate multiple molecular configurations consistent with each latent space location. We train the cWGAN over $(\mathbf{x}^\gamma,\boldsymbol{\psi}^\gamma)$ pairs by encoding each frame $\gamma$ of the MD training trajectory into the latent space using the SRV. The trained cWGAN decoder $D$ (Eqn.~\ref{eqn:scheme}) generates molecular configurations from the latent space trajectory produced by the propagator (Fig.~\ref{lss_diagram}).

\section{Results and Discussion}

\subsection{4-well potential}

We validate the LSS in an application to a 1D four-well potential \cite{Prinz2011} $V(x) = 2(x^8 + 0.8 e^{-80x^2} + 0.2 e^{-80(x-0.5)^2} + 0.5 e^{-40 (x + 0.5)^2} )$ for which analytical solutions are available. In this simple 1D system we construct the propagator directly in $x = \psi \in \mathbb{R}^1$, so encoding and decoding are unnecessary and this test validates that the MDN can learn transition densities $p_\tau(x_{t+\tau} | x_t)$ to accurately reproduce the system thermodynamics and kinetics. We generate a $5 \times 10^6$ time step Brownian dynamics trajectory in a dimensionless gauge with diffusivity $D$=$k_B T$=1000 and a time step $\Delta t$=0.001 \cite{beauchamp2011msmbuilder2}. A MDN was trained using Adam \cite{kingma2014adam} with early stopping over the $[0,1]$ scaled trajectory at a lag time of $\tau$=100, with $C$=8 Gaussian kernels, and two hidden layers of 100 neurons with ReLU activations \cite{goodfellow2016deep}. The trained MDN was used to generate a $5 \times 10^4$ step trajectory of the same length as the training data. Analytical transition densities were computed by partitioning the domain into 100 equal bins and defining the probability of moving from bin $i$ to bin $j$ as $p(j|i) = C_i e^{-(V_j-V_i)}$ for $|i-j| \leq 1$, where $V_i$ is the potential at the center of bin $i$ and $C_i$ normalizes the total transition probability of bin $i$ \cite{Chen2019a}.

\begin{figure}[ht!]
\begin{centering}
  \includegraphics[width=\linewidth]{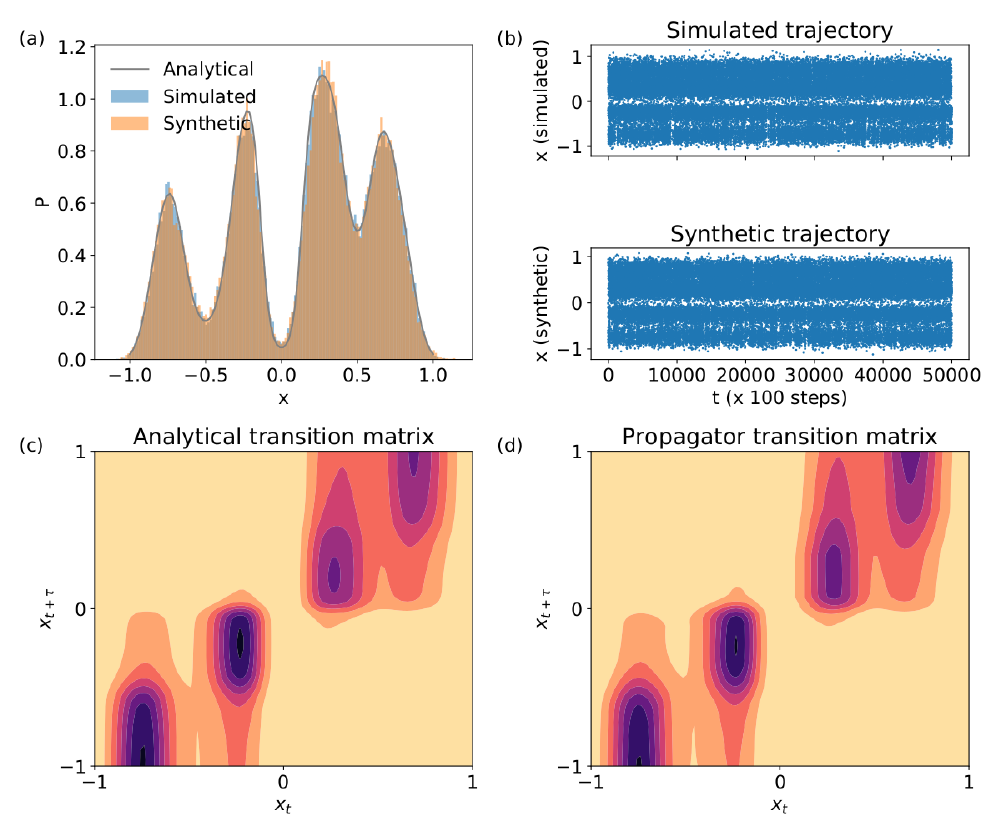}
  \caption{
  Validation of the LSS in a 1D four-well potential. The MDN propagator predicts (a) a stationary distribution, (b) kinetic transitions, and (c,d) transition densities in excellent accord with analytical and Brownian dynamics results.
  }
  \label{four_well_panel}
\end{centering}
\end{figure}

The Brownian dynamics and synthetic LSS stationary distributions are in quantitative agreement with the analytical solution for the stationary density (Fig.~\ref{four_well_panel}a) and show very similar kinetic behaviors in their transitions between the four metastable wells (Fig.~\ref{four_well_panel}b). This agreement is due to the excellent correspondence between the analytical and learned transition densities (Fig.~\ref{four_well_panel}c,d).

\subsection{Trp-cage miniprotein}

We train our LSS over the 208 $\mu$s all-atom simulation of the 20-residue TC10b K8A mutant of the Trp-cage mini-protein performed by D.E.~Shaw Research \cite{Lindorff-Larsen2011}. Generation of these MD trajectories would require $\sim$2.5 days (2 million CPU-h) on the special purpose Anton-2 supercomputer or $\sim$6 months on a commodity GPU card \cite{Shaw2014}.

The SRV encoder was trained over a featurization the trajectory employing backbone and sidechain torsions and C$\alpha$ pairwise distances as informative and roto-translationally invariant descriptors \cite{Sidky2019}. We trained a SRV with two hidden layers with 100 neurons, $\tanh$ activations, and batch normalization using Adam \cite{kingma2014adam} with a batch size of 200,000, learning rate of 0.01, and early stopping based on the validation VAMP-2 score \cite{Chen2019a,Sidky2019}. A lag-time of $\tau$=20 ns was chosen based on convergence of the transfer operator eigenvalues, and a $m$=3-dimensional latent space encoding based on a gap in the eigenvalue spectrum. The MDN propagator was trained over the latent space projection of the MD trajectory at a lag time of $\tau$=20 ns using Adam \cite{kingma2014adam} with early stopping, $C$=24 Gaussian kernels, and two hidden layers of 100 neurons with ReLU activations. The cWGAN decoder comprised a generator and critic with three hidden layers of 200 neurons with Swish \cite{ramachandran2017swish} activations and a $d$=50-dimensional noise vector. The training loss stabilized after 52 epochs. The cWGAN is trained to generate the Trp-cage C$\alpha$ backbone by roto-translationally aligning MD training configurations to a reference structure. Training of the full LSS pipeline required $\sim$1 GPU-h on a NVIDIA GeForce GTX 1080 Ti GPU core.

The trained LSS was used to produce 100$\times$208 $\mu$s synthetic trajectories each requiring $\sim$5 s on a single NVIDIA GeForce GTX 1080 Ti GPU core. The LSS trajectories comprise the same total number of frames as the 208 $\mu$s all-atom trajectory but contain $\sim$1070 folding/unfolding transitions compared to just 12 in the training data and were generated at six orders of magnitude lower cost. This observation illuminates the crux of the value of the approach: the LSS learns a kinetic model over limited MD training data and is then used to generate vastly longer novel all-atom trajectories that enable the observation of states and events that are only sparsely sampled in the training data. We now validate the thermodynamic, structural, and kinetic predictions of the LSS.

\begin{figure}[ht!]
\begin{centering}
  \includegraphics[width=0.8\linewidth]{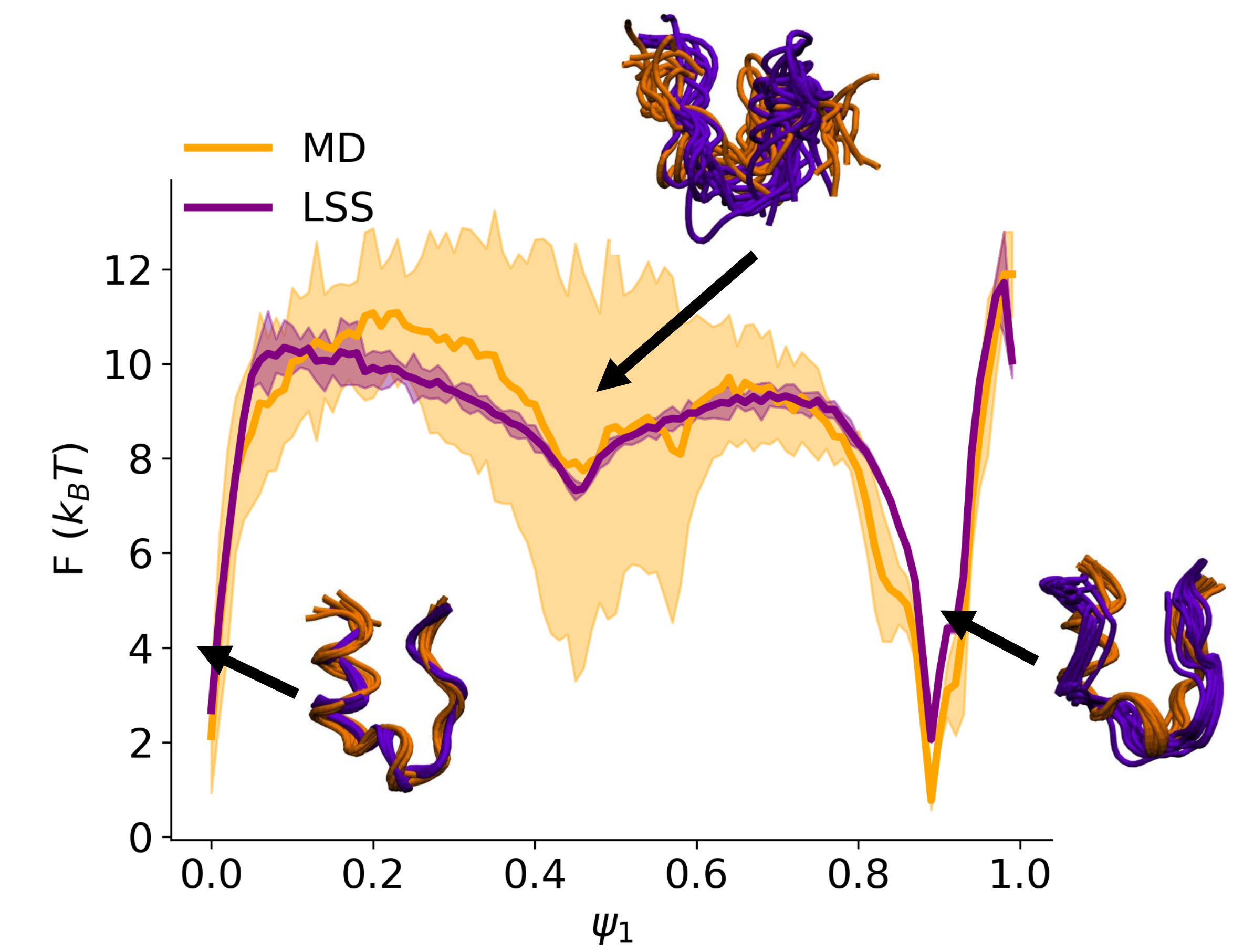}
  \caption{
Free energy profiles for the MD and LSS trajectories projected into the slowest latent space coordinate $\psi_1$. Shaded backgrounds represent standard errors estimated by five-fold block averaging.  The profiles agree within a 0.91 $k_B T$ root mean squared error. Ten representative structures from the MD and LSS ensembles are sampled from the folded ($\psi_1$$\approx$0), unfolded ($\psi_1$$\approx$0.9), and metastable ($\psi_1$$\approx$0.45) regions. 
  }
  \label{fes_trpcage}
\end{centering}
\end{figure}

\textbf{Thermodynamics.} The free energy profiles projected into the slowest latent space coordinate $F(\psi_1)$=$-k_B T \ln \left( q(\psi_1) \right)$ show excellent correspondence between the MD and LSS (Fig.~\ref{fes_trpcage}). The free energy of the folded ($\psi_1$$\approx$0) and unfolded ($\psi_1$$\approx$0.9) basins and transition barrier are in quantitative agreement with a root mean squared error between the aligned profiles of 0.91 $k_B T$. The LSS profiles contain 10-fold lower statistical uncertainties than the MD over the same number of frames due to the 100-fold longer LSS data set enabled by their exceedingly low computational cost.

\textbf{Structures.} The MD and LSS molecular structures within the folded basin ($\psi_1\approx 0$) and metastable transition state ($\psi_1\approx 0.45$) possess a relative $C_\alpha$-RMSD of 0.29 nm and 0.37 nm, respectively. Relative to the Trp-cage native state (PDB ID: 2JOF), the MD and LSS folded configurations possess a $C_\alpha$-RMSD of 0.20 nm and 0.28 nm, respectively. The mean and standard deviation of the time-averaged radii of gyration (\blauw{$R_g$}) for the MD (0.87$\pm$0.16) nm and LSS (0.87$\pm$0.13) nm trajectories are indistinguishable with standard errors computed by five-fold block averaging. These results demonstrate that the LSS molecular structures are in excellent accord with MD.

\textbf{Kinetics.} We compare the MD and LSS kinetics through the autocorrelation times corresponding to the relaxation time scales associated with the $m$=3 leading kinetic processes. All three time scales are in excellent agreement and again the LSS uncertainties are approximately 10-fold lower than the MD (Table~\ref{table:timescales}).

\begin{table}[ht!]
  \caption{Implied time scales of leading Trp-cage modes. Standard errors estimated by five-fold block averaging.}
  \begin{tabular*}{0.48\textwidth}{@{\extracolsep{\fill}}ccc}
  Timescale & \textrm{MD ($\mu$s)} & \textrm{LSS ($\mu$s)} \\
  \hline
  $t_1$ & 3.00 $\pm$ 0.61 & 2.89 $\pm$ 0.12 \\ 
  $t_2$ & 0.54 $\pm$ 0.37 & 0.43 $\pm$ 0.04 \\ 
  $t_3$ & 0.45 $\pm$ 0.12 & 0.42 $\pm$ 0.01 \\ 
  \end{tabular*}
\label{table:timescales}
\end{table}

\begin{figure}[ht!]
\begin{centering}
  \includegraphics[width=0.9\linewidth]{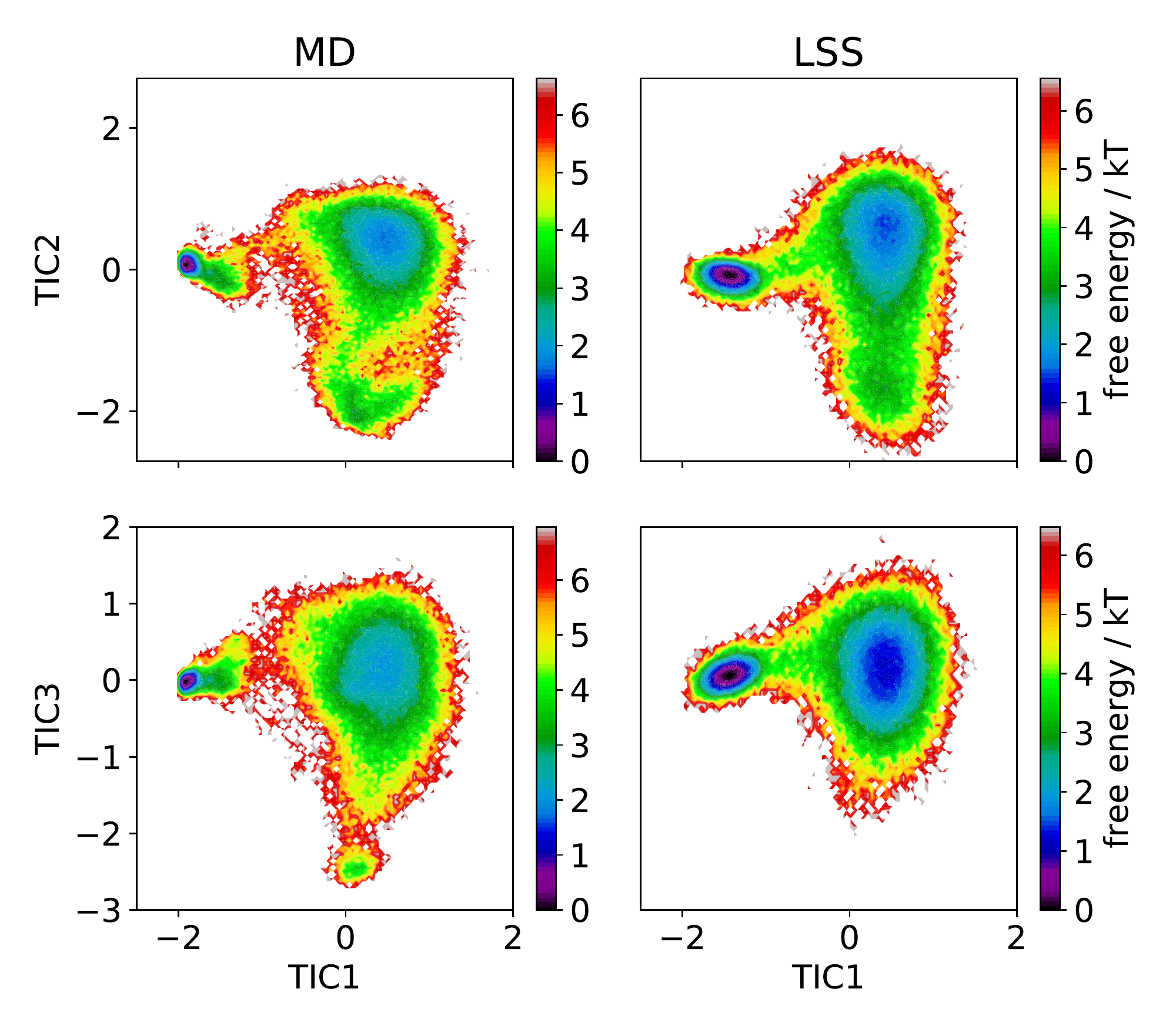}
  \caption{
  Free energy profiles of the MD and LSS trajectories projected into the leading three MD TICA coordinates.
  }
  \label{tica_trpcage}
\end{centering}
\end{figure}

We then employ time-lagged independent component analysis (\blauw{TICA}) \cite{perez2013identification, noe2013variational, nuske2014variational, noe2015kinetic, noe2016commute,
perez2016hierarchical, schwantes2013improvements, klus2018data} to determine whether the LSS trajectory possesses the same slow (linear) subspace as the MD. We featurize the trajectories with pairwise C$\alpha$ distances and perform TICA at a lag time of $\tau$=20 ns. Projection of the free energy surfaces into the leading three MD TICA coordinates show that the leading kinetic variance in the MD data is quite accurately reproduced by the LSS (Fig.~\ref{tica_trpcage}). The only substantive disagreement is absence in the LSS projection of a small high-free energy metastable state at (TIC1 $\approx$ $0$, TIC3 $\approx$ $-2.5$) corresponding to configurations with Pro18 dihedral angles $\psi\approx (-50)^\circ$. These configurations are only transiently occupied due to rare Pro18 dihedral flips that occur only twice during the 208 $\mu$s MD trajectory and are not contained in the $m$=3-dimensional latent space.

\section{Conclusions}

We have presented LSS as a method to learn efficient kinetic models by training three state-of-the-art deep learning networks over MD training data and then using the trained model to generate novel all-atom trajectories at six orders of magnitude lower cost. The spirit of the approach is similar to MSM-based and equation-free approaches that use limited MD training data to parameterize highly-efficient kinetic models that can then be used to generate dynamical trajectories over vastly longer time scales than are possible with conventional MD. In contrast to these approaches, the absence of any discretization of the configurational space and provisioning with a molecular decoder enables the LSS to produce continuous all-atom molecular trajectories. Importantly, the probabilistic and generative nature of the approach means that the generated molecular trajectories are novel and not simply a reproduction of the training data, and the statistics of these trajectories accurately reproduce the structural, thermodynamic and kinetic properties of the molecular system.

The dramatic reduction in the cost of trajectory generation opens a host of valuable possibilities: vastly improved sampling of configurational space and dynamical transitions enable estimates of thermodynamic averages and kinetic rates with exceedingly low statistical uncertainties; parameterization of kinetic models with modest training data enable the production of ultra-long trajectories on commodity computing hardware; representation of the kinetic model as the parameters of a trio of deep networks enables efficient sharing of a ``simulator in a box'' that can then be used for rapid on-demand trajectory generation. The properties of the trained kinetic model -- the dimensionality of the slow latent space, the structural correspondence of the slow modes, and the transition probabilities of the propagator -- also provide fundamental insight and understanding of the physical properties of the molecular system. 

As with all data-driven approaches, the primary deficiency of the LSS approach is that the resulting kinetic models are not necessarily transferable to other conditions or systems and are subject to systematic errors due to approximations in the molecular force fields and incomplete sampling of the relevant configurational space in the training data. The latter issue means that although the generated LSS trajectories are -- similar to MSM-based and equation-free approaches -- largely interpolative. The stochastic nature of the MDN propagator and generative nature of the cWGAN generator means that we may anticipate local extrapolations beyond the exact training configurations \cite{Wu2018}. There is no expectation, however, that the trained model will discover new metastable states or kinetic transitions, and certainly not do so with the correct thermodynamic weights or dynamical time scales. 

The present work has demonstrated LSS in a data-rich training regime where the MD training data comprehensively samples configurational space. The next step is to establish an adaptive sampling paradigm -- similar to that in MSM construction \cite{pande2010everything} and some enhanced sampling techniques \cite{chen2018collective,chiavazzo2017intrinsic,preto2014fast,zheng2013rapid} -- to enable its application in a data-poor regime. The adaptive sampling approach interrogates the kinetic model to identify under-sampled states and transitions that contribute most to uncertainties in the model predictions (i.e., ``known unknowns'') and initializes new MD simulations to collect additional training data in these regions. This interleaving of MD training data collection and model retraining can dramatically reduce the required quantity of training data \cite{pande2010everything}. Moreover, new simulations initialized in under-sampled regions may also occasionally be expected to transition into new configurational states not present in the initial training data (i.e., ``unknown unknowns'') \cite{preto2014fast}. Iterating this process until convergence can expand the range of the trained kinetic model to encompass the relevant configurational space and minimize the cost of training data collection. 

Finally, we also envisage applications of the LSS approach to other fields of dynamical modeling where stiff or multi-scale systems of ordinary or partial differential equations, or the presence of activated processes or rare events introduces a separation of time scales between the integration time step and events of interest. For example, there may be profitable adaptations of the approach for dynamical modeling in such fields as cosmology, ecology, immunology, epidemiology, and climatology.

\section*{Conflicts of interest}

A.L.F. is a consultant of Evozyne and a co-author of US Provisional Patents 62/853,919 and 62/900,420 and International Patent Application PCT/US2020/035206.


\section*{Acknowledgements}

This work was supported by MICCoM (Midwest Center for Computational Materials), as part of the Computational Materials Science Program funded by the U.S. Department of Energy, Office of Science, Basic Energy Sciences, Materials Sciences and Engineering Division and the National Science Foundation under Grant No.~CHE-1841805. H.S.~acknowledges support from the National Science Foundation Molecular Software Sciences Institute (MolSSI) Software Fellows program (Grant No.~ACI-1547580) \cite{krylov2018perspective,wilkins2018nsf}. We are grateful to D.E.~Shaw Research for sharing the Trp-cage simulation trajectories.



\balance


\bibliography{references} 

\providecommand*{\mcitethebibliography}{\thebibliography}
\csname @ifundefined\endcsname{endmcitethebibliography}
{\let\endmcitethebibliography\endthebibliography}{}
\begin{mcitethebibliography}{73}
\providecommand*{\natexlab}[1]{#1}
\providecommand*{\mciteSetBstSublistMode}[1]{}
\providecommand*{\mciteSetBstMaxWidthForm}[2]{}
\providecommand*{\mciteBstWouldAddEndPuncttrue}
  {\def\EndOfBibitem{\unskip.}}
\providecommand*{\mciteBstWouldAddEndPunctfalse}
  {\let\EndOfBibitem\relax}
\providecommand*{\mciteSetBstMidEndSepPunct}[3]{}
\providecommand*{\mciteSetBstSublistLabelBeginEnd}[3]{}
\providecommand*{\EndOfBibitem}{}
\mciteSetBstSublistMode{f}
\mciteSetBstMaxWidthForm{subitem}
{(\emph{\alph{mcitesubitemcount}})}
\mciteSetBstSublistLabelBeginEnd{\mcitemaxwidthsubitemform\space}
{\relax}{\relax}

\bibitem[Frenkel and Smit(2002)]{frenkel2002understanding}
D.~Frenkel and B.~Smit, \emph{Understanding Molecular Simulation: From
  algorithms to applications}, Academic Press, San Diego, 2002\relax
\mciteBstWouldAddEndPuncttrue
\mciteSetBstMidEndSepPunct{\mcitedefaultmidpunct}
{\mcitedefaultendpunct}{\mcitedefaultseppunct}\relax
\EndOfBibitem
\bibitem[Stone \emph{et~al.}(2010)Stone, Hardy, Ufimtsev, and
  Schulten]{stone2010gpu}
J.~E. Stone, D.~J. Hardy, I.~S. Ufimtsev and K.~Schulten, \emph{Journal of
  Molecular Graphics and Modelling}, 2010, \textbf{29}, 116--125\relax
\mciteBstWouldAddEndPuncttrue
\mciteSetBstMidEndSepPunct{\mcitedefaultmidpunct}
{\mcitedefaultendpunct}{\mcitedefaultseppunct}\relax
\EndOfBibitem
\bibitem[Shaw \emph{et~al.}(2014)Shaw, Grossman, Bank, Batson, Butts, Chao,
  Deneroff, Dror, Even, Fenton, Forte, Gagliardo, Gill, Greskamp, Ho, Ierardi,
  Iserovich, Kuskin, Larson, Layman, Lee, Lerer, Li, Killebrew, Mackenzie, Mok,
  Moraes, Mueller, Nociolo, Peticolas, Quan, Ramot, Salmon, Scarpazza, {Ben
  Schafer}, Siddique, Snyder, Spengler, Tang, Theobald, Toma, Towles, Vitale,
  Wang, and Young]{Shaw2014}
D.~E. Shaw, J.~P. Grossman, J.~A. Bank, B.~Batson, J.~A. Butts, J.~C. Chao,
  M.~M. Deneroff, R.~O. Dror, A.~Even, C.~H. Fenton, A.~Forte, J.~Gagliardo,
  G.~Gill, B.~Greskamp, C.~R. Ho, D.~J. Ierardi, L.~Iserovich, J.~S. Kuskin,
  R.~H. Larson, T.~Layman, L.~S. Lee, A.~K. Lerer, C.~Li, D.~Killebrew, K.~M.
  Mackenzie, S.~Y.~H. Mok, M.~A. Moraes, R.~Mueller, L.~J. Nociolo, J.~L.
  Peticolas, T.~Quan, D.~Ramot, J.~K. Salmon, D.~P. Scarpazza, U.~{Ben
  Schafer}, N.~Siddique, C.~W. Snyder, J.~Spengler, P.~T.~P. Tang, M.~Theobald,
  H.~Toma, B.~Towles, B.~Vitale, S.~C. Wang and C.~Young, SC '14: Proceedings
  of the International Conference for High Performance Computing, Networking,
  Storage and Analysis, 2014, pp. 41--53\relax
\mciteBstWouldAddEndPuncttrue
\mciteSetBstMidEndSepPunct{\mcitedefaultmidpunct}
{\mcitedefaultendpunct}{\mcitedefaultseppunct}\relax
\EndOfBibitem
\bibitem[Phillips \emph{et~al.}(2005)Phillips, Braun, Wang, Gumbart,
  Tajkhorshid, Villa, Chipot, Skeel, Kale, and Schulten]{phillips2005scalable}
J.~C. Phillips, R.~Braun, W.~Wang, J.~Gumbart, E.~Tajkhorshid, E.~Villa,
  C.~Chipot, R.~D. Skeel, L.~Kale and K.~Schulten, \emph{Journal of
  Computational Chemistry}, 2005, \textbf{26}, 1781--1802\relax
\mciteBstWouldAddEndPuncttrue
\mciteSetBstMidEndSepPunct{\mcitedefaultmidpunct}
{\mcitedefaultendpunct}{\mcitedefaultseppunct}\relax
\EndOfBibitem
\bibitem[Chow \emph{et~al.}(2008)Chow, Rendleman, Bowers, Dror, Hughes,
  Gullingsrud, Sacerdoti, and Shaw]{chow2008desmond}
E.~Chow, C.~A. Rendleman, K.~J. Bowers, R.~O. Dror, D.~H. Hughes,
  J.~Gullingsrud, F.~D. Sacerdoti and D.~E. Shaw, \emph{Desmond performance on
  a cluster of multicore processors (DESRES/TR--2008-01)}, {DE Shaw Research}
  technical report, 2008\relax
\mciteBstWouldAddEndPuncttrue
\mciteSetBstMidEndSepPunct{\mcitedefaultmidpunct}
{\mcitedefaultendpunct}{\mcitedefaultseppunct}\relax
\EndOfBibitem
\bibitem[Glaser \emph{et~al.}(2015)Glaser, Nguyen, Anderson, Lui, Spiga,
  Millan, Morse, and Glotzer]{glaser2015strong}
J.~Glaser, T.~D. Nguyen, J.~A. Anderson, P.~Lui, F.~Spiga, J.~A. Millan, D.~C.
  Morse and S.~C. Glotzer, \emph{Computer Physics Communications}, 2015,
  \textbf{192}, 97--107\relax
\mciteBstWouldAddEndPuncttrue
\mciteSetBstMidEndSepPunct{\mcitedefaultmidpunct}
{\mcitedefaultendpunct}{\mcitedefaultseppunct}\relax
\EndOfBibitem
\bibitem[Plimpton(1993)]{plimpton1993fast}
S.~Plimpton, \emph{Fast parallel algorithms for short-range molecular dynamics
  (SAND-91-1144)}, {Sandia National Laboratory} technical report, 1993\relax
\mciteBstWouldAddEndPuncttrue
\mciteSetBstMidEndSepPunct{\mcitedefaultmidpunct}
{\mcitedefaultendpunct}{\mcitedefaultseppunct}\relax
\EndOfBibitem
\bibitem[Abraham \emph{et~al.}(2002)Abraham, Walkup, Gao, Duchaineau,
  De~La~Rubia, and Seager]{abraham2002work}
F.~F. Abraham, R.~Walkup, H.~Gao, M.~Duchaineau, T.~D. De~La~Rubia and
  M.~Seager, \emph{Proceedings of the National Academy of Sciences of the
  United States of America}, 2002, \textbf{99}, 5783--5787\relax
\mciteBstWouldAddEndPuncttrue
\mciteSetBstMidEndSepPunct{\mcitedefaultmidpunct}
{\mcitedefaultendpunct}{\mcitedefaultseppunct}\relax
\EndOfBibitem
\bibitem[Abraham \emph{et~al.}(2002)Abraham, Walkup, Gao, Duchaineau,
  De~La~Rubia, and Seager]{abraham2002brittle}
F.~F. Abraham, R.~Walkup, H.~Gao, M.~Duchaineau, T.~D. De~La~Rubia and
  M.~Seager, \emph{Proceedings of the National Academy of Sciences of the
  United States of America}, 2002, \textbf{99}, 5777--5782\relax
\mciteBstWouldAddEndPuncttrue
\mciteSetBstMidEndSepPunct{\mcitedefaultmidpunct}
{\mcitedefaultendpunct}{\mcitedefaultseppunct}\relax
\EndOfBibitem
\bibitem[Tchipev \emph{et~al.}(2019)Tchipev, Seckler, Heinen, Vrabec, Gratl,
  Horsch, Bernreuther, Glass, Niethammer, Hammer, Krischok, Resch,
  Kranzlm{\"u}ller, Hasse, Bungartz, and Neumann]{tchipev2019twetris}
N.~Tchipev, S.~Seckler, M.~Heinen, J.~Vrabec, F.~Gratl, M.~Horsch,
  M.~Bernreuther, C.~W. Glass, C.~Niethammer, N.~Hammer, B.~Krischok, M.~Resch,
  D.~Kranzlm{\"u}ller, H.~Hasse, H.-J. Bungartz and P.~Neumann, \emph{The
  International Journal of High Performance Computing Applications}, 2019,
  \textbf{33}, 838--854\relax
\mciteBstWouldAddEndPuncttrue
\mciteSetBstMidEndSepPunct{\mcitedefaultmidpunct}
{\mcitedefaultendpunct}{\mcitedefaultseppunct}\relax
\EndOfBibitem
\bibitem[Elber(2016)]{elber2016perspective}
R.~Elber, \emph{The Journal of Chemical Physics}, 2016, \textbf{144},
  060901\relax
\mciteBstWouldAddEndPuncttrue
\mciteSetBstMidEndSepPunct{\mcitedefaultmidpunct}
{\mcitedefaultendpunct}{\mcitedefaultseppunct}\relax
\EndOfBibitem
\bibitem[Torrie and Valleau(1977)]{torrie1977nonphysical}
G.~M. Torrie and J.~P. Valleau, \emph{Journal of Computational Physics}, 1977,
  \textbf{23}, 187--199\relax
\mciteBstWouldAddEndPuncttrue
\mciteSetBstMidEndSepPunct{\mcitedefaultmidpunct}
{\mcitedefaultendpunct}{\mcitedefaultseppunct}\relax
\EndOfBibitem
\bibitem[McDonald and Singer(1967)]{mcdonald1967machine}
I.~R. McDonald and K.~Singer, \emph{The Journal of Chemical Physics}, 1967,
  \textbf{47}, 4766--4772\relax
\mciteBstWouldAddEndPuncttrue
\mciteSetBstMidEndSepPunct{\mcitedefaultmidpunct}
{\mcitedefaultendpunct}{\mcitedefaultseppunct}\relax
\EndOfBibitem
\bibitem[Abrams and Bussi(2014)]{abrams2014enhanced}
C.~Abrams and G.~Bussi, \emph{Entropy}, 2014, \textbf{16}, 163--199\relax
\mciteBstWouldAddEndPuncttrue
\mciteSetBstMidEndSepPunct{\mcitedefaultmidpunct}
{\mcitedefaultendpunct}{\mcitedefaultseppunct}\relax
\EndOfBibitem
\bibitem[Miao and McCammon(2016)]{miao2016unconstrained}
Y.~Miao and J.~A. McCammon, \emph{Molecular Simulation}, 2016, \textbf{42},
  1046--1055\relax
\mciteBstWouldAddEndPuncttrue
\mciteSetBstMidEndSepPunct{\mcitedefaultmidpunct}
{\mcitedefaultendpunct}{\mcitedefaultseppunct}\relax
\EndOfBibitem
\bibitem[Sidky \emph{et~al.}(2020)Sidky, Chen, and Ferguson]{sidky2020machine}
H.~Sidky, W.~Chen and A.~L. Ferguson, \emph{Molecular Physics}, 2020,
  \textbf{118}, 1--21\relax
\mciteBstWouldAddEndPuncttrue
\mciteSetBstMidEndSepPunct{\mcitedefaultmidpunct}
{\mcitedefaultendpunct}{\mcitedefaultseppunct}\relax
\EndOfBibitem
\bibitem[Chodera \emph{et~al.}(2011)Chodera, Swope, No{\'e}, Prinz, Shirts, and
  Pande]{chodera2011dynamical}
J.~D. Chodera, W.~C. Swope, F.~No{\'e}, J.-H. Prinz, M.~R. Shirts and V.~S.
  Pande, \emph{The Journal of Chemical Physics}, 2011, \textbf{134},
  06B612\relax
\mciteBstWouldAddEndPuncttrue
\mciteSetBstMidEndSepPunct{\mcitedefaultmidpunct}
{\mcitedefaultendpunct}{\mcitedefaultseppunct}\relax
\EndOfBibitem
\bibitem[Donati and Keller(2018)]{donati2018girsanov}
L.~Donati and B.~G. Keller, \emph{The Journal of Chemical Physics}, 2018,
  \textbf{149}, 072335\relax
\mciteBstWouldAddEndPuncttrue
\mciteSetBstMidEndSepPunct{\mcitedefaultmidpunct}
{\mcitedefaultendpunct}{\mcitedefaultseppunct}\relax
\EndOfBibitem
\bibitem[No{\'e}(2018)]{noe2018machine}
F.~No{\'e}, \emph{arXiv preprint arXiv:1812.07669}, 2018,
  https://arxiv.org/abs/1812.07669v1\relax
\mciteBstWouldAddEndPuncttrue
\mciteSetBstMidEndSepPunct{\mcitedefaultmidpunct}
{\mcitedefaultendpunct}{\mcitedefaultseppunct}\relax
\EndOfBibitem
\bibitem[Fern{\'{a}}ndez(2020)]{Fernandez2020}
A.~Fern{\'{a}}ndez, \emph{Annalen der Physik}, 2020, \textbf{532}, 1--5\relax
\mciteBstWouldAddEndPuncttrue
\mciteSetBstMidEndSepPunct{\mcitedefaultmidpunct}
{\mcitedefaultendpunct}{\mcitedefaultseppunct}\relax
\EndOfBibitem
\bibitem[Husic and Pande(2018)]{husic2018markov}
B.~E. Husic and V.~S. Pande, \emph{Journal of the American Chemical Society},
  2018, \textbf{140}, 2386--2396\relax
\mciteBstWouldAddEndPuncttrue
\mciteSetBstMidEndSepPunct{\mcitedefaultmidpunct}
{\mcitedefaultendpunct}{\mcitedefaultseppunct}\relax
\EndOfBibitem
\bibitem[Pande \emph{et~al.}(2010)Pande, Beauchamp, and
  Bowman]{pande2010everything}
V.~S. Pande, K.~Beauchamp and G.~R. Bowman, \emph{Methods}, 2010, \textbf{52},
  99--105\relax
\mciteBstWouldAddEndPuncttrue
\mciteSetBstMidEndSepPunct{\mcitedefaultmidpunct}
{\mcitedefaultendpunct}{\mcitedefaultseppunct}\relax
\EndOfBibitem
\bibitem[Prinz \emph{et~al.}(2011)Prinz, Wu, Sarich, Keller, Senne, Held,
  Chodera, Schtte, and No{\'{e}}]{Prinz2011}
J.~H. Prinz, H.~Wu, M.~Sarich, B.~Keller, M.~Senne, M.~Held, J.~D. Chodera,
  C.~Schtte and F.~No{\'{e}}, \emph{Journal of Chemical Physics}, 2011,
  \textbf{134}, 174105\relax
\mciteBstWouldAddEndPuncttrue
\mciteSetBstMidEndSepPunct{\mcitedefaultmidpunct}
{\mcitedefaultendpunct}{\mcitedefaultseppunct}\relax
\EndOfBibitem
\bibitem[Bowman \emph{et~al.}(2013)Bowman, Pande, and
  No{\'e}]{bowman2013introduction}
G.~R. Bowman, V.~S. Pande and F.~No{\'e}, \emph{An Introduction to Markov State
  Models and Their Application to Long Timescale Molecular Simulation},
  Springer Science \& Business Media, 2013, vol. 797\relax
\mciteBstWouldAddEndPuncttrue
\mciteSetBstMidEndSepPunct{\mcitedefaultmidpunct}
{\mcitedefaultendpunct}{\mcitedefaultseppunct}\relax
\EndOfBibitem
\bibitem[Sidky \emph{et~al.}(2019)Sidky, Chen, and Ferguson]{Sidky2019}
H.~Sidky, W.~Chen and A.~L. Ferguson, \emph{Journal of Physical Chemistry B},
  2019, \textbf{123}, 7999--8009\relax
\mciteBstWouldAddEndPuncttrue
\mciteSetBstMidEndSepPunct{\mcitedefaultmidpunct}
{\mcitedefaultendpunct}{\mcitedefaultseppunct}\relax
\EndOfBibitem
\bibitem[Wehmeyer \emph{et~al.}(2019)Wehmeyer, Scherer, Hempel, Husic, Olsson,
  and No{\'{e}}]{Wehmeyer2019}
C.~Wehmeyer, M.~K. Scherer, T.~Hempel, B.~E. Husic, S.~Olsson and F.~No{\'{e}},
  \emph{Living Journal of Computational Molecular Science}, 2019, \textbf{1},
  1--12\relax
\mciteBstWouldAddEndPuncttrue
\mciteSetBstMidEndSepPunct{\mcitedefaultmidpunct}
{\mcitedefaultendpunct}{\mcitedefaultseppunct}\relax
\EndOfBibitem
\bibitem[Mardt \emph{et~al.}(2018)Mardt, Pasquali, Wu, and
  No{\'{e}}]{Mardt2018}
A.~Mardt, L.~Pasquali, H.~Wu and F.~No{\'{e}}, \emph{Nature Communications},
  2018, \textbf{9}, 5\relax
\mciteBstWouldAddEndPuncttrue
\mciteSetBstMidEndSepPunct{\mcitedefaultmidpunct}
{\mcitedefaultendpunct}{\mcitedefaultseppunct}\relax
\EndOfBibitem
\bibitem[Wu and No{\'e}(2020)]{wu2017variational}
H.~Wu and F.~No{\'e}, \emph{Journal of Nonlinear Science}, 2020, \textbf{30},
  23--66\relax
\mciteBstWouldAddEndPuncttrue
\mciteSetBstMidEndSepPunct{\mcitedefaultmidpunct}
{\mcitedefaultendpunct}{\mcitedefaultseppunct}\relax
\EndOfBibitem
\bibitem[Kevrekidis \emph{et~al.}(2003)Kevrekidis, Gear, Hyman, Kevrekidis,
  Runborg, and Theodoropoulos]{kevrekidis2003equation}
I.~G. Kevrekidis, C.~W. Gear, J.~M. Hyman, P.~G. Kevrekidis, O.~Runborg and
  C.~Theodoropoulos, \emph{Communications in Mathematical Sciences}, 2003,
  \textbf{1}, 715--762\relax
\mciteBstWouldAddEndPuncttrue
\mciteSetBstMidEndSepPunct{\mcitedefaultmidpunct}
{\mcitedefaultendpunct}{\mcitedefaultseppunct}\relax
\EndOfBibitem
\bibitem[Kevrekidis \emph{et~al.}(2004)Kevrekidis, Gear, and
  Hummer]{kevrekidis2004equation}
I.~G. Kevrekidis, C.~W. Gear and G.~Hummer, \emph{AIChE Journal}, 2004,
  \textbf{50}, 1346--1355\relax
\mciteBstWouldAddEndPuncttrue
\mciteSetBstMidEndSepPunct{\mcitedefaultmidpunct}
{\mcitedefaultendpunct}{\mcitedefaultseppunct}\relax
\EndOfBibitem
\bibitem[Kevrekidis and Samaey(2009)]{kevrekidis2009equation}
I.~G. Kevrekidis and G.~Samaey, \emph{Annual Review of Physical Chemistry},
  2009, \textbf{60}, 321--344\relax
\mciteBstWouldAddEndPuncttrue
\mciteSetBstMidEndSepPunct{\mcitedefaultmidpunct}
{\mcitedefaultendpunct}{\mcitedefaultseppunct}\relax
\EndOfBibitem
\bibitem[Mori(1965)]{Mori1965}
H.~Mori, \emph{Progress of Theoretical Physics}, 1965, \textbf{33},
  423--455\relax
\mciteBstWouldAddEndPuncttrue
\mciteSetBstMidEndSepPunct{\mcitedefaultmidpunct}
{\mcitedefaultendpunct}{\mcitedefaultseppunct}\relax
\EndOfBibitem
\bibitem[Zwanzig(1973)]{Zwanzig1973}
R.~Zwanzig, \emph{Journal of Statistical Physics}, 1973, \textbf{9},
  215--220\relax
\mciteBstWouldAddEndPuncttrue
\mciteSetBstMidEndSepPunct{\mcitedefaultmidpunct}
{\mcitedefaultendpunct}{\mcitedefaultseppunct}\relax
\EndOfBibitem
\bibitem[Zwanzig(2001)]{zwanzig2001nonequilibrium}
R.~Zwanzig, \emph{Nonequilibrium Statistical Mechanics}, Oxford University
  Press, Oxford, 2001\relax
\mciteBstWouldAddEndPuncttrue
\mciteSetBstMidEndSepPunct{\mcitedefaultmidpunct}
{\mcitedefaultendpunct}{\mcitedefaultseppunct}\relax
\EndOfBibitem
\bibitem[Risken and Frank(2012)]{risken2012fokker}
H.~Risken and T.~Frank, \emph{The Fokker-Planck Equation: Methods of Solution
  and Applications}, Springer Verlag, Berlin Heidelberg New York, 2nd edn,
  2012\relax
\mciteBstWouldAddEndPuncttrue
\mciteSetBstMidEndSepPunct{\mcitedefaultmidpunct}
{\mcitedefaultendpunct}{\mcitedefaultseppunct}\relax
\EndOfBibitem
\bibitem[Wehmeyer and No{\'{e}}(2018)]{Wehmeyer2018}
C.~Wehmeyer and F.~No{\'{e}}, \emph{Journal of Chemical Physics}, 2018,
  \textbf{148}, 241703\relax
\mciteBstWouldAddEndPuncttrue
\mciteSetBstMidEndSepPunct{\mcitedefaultmidpunct}
{\mcitedefaultendpunct}{\mcitedefaultseppunct}\relax
\EndOfBibitem
\bibitem[Hern{\'{a}}ndez \emph{et~al.}(2018)Hern{\'{a}}ndez, Wayment-Steele,
  Sultan, Husic, and Pande]{Hernandez2017}
C.~X. Hern{\'{a}}ndez, H.~K. Wayment-Steele, M.~M. Sultan, B.~E. Husic and
  V.~S. Pande, \emph{Physical Review E}, 2018, \textbf{97}, 1--12\relax
\mciteBstWouldAddEndPuncttrue
\mciteSetBstMidEndSepPunct{\mcitedefaultmidpunct}
{\mcitedefaultendpunct}{\mcitedefaultseppunct}\relax
\EndOfBibitem
\bibitem[Lusch \emph{et~al.}(2018)Lusch, Kutz, and Brunton]{Lusch2018}
B.~Lusch, J.~N. Kutz and S.~L. Brunton, \emph{Nature Communications}, 2018,
  \textbf{9}, 4950\relax
\mciteBstWouldAddEndPuncttrue
\mciteSetBstMidEndSepPunct{\mcitedefaultmidpunct}
{\mcitedefaultendpunct}{\mcitedefaultseppunct}\relax
\EndOfBibitem
\bibitem[Chen \emph{et~al.}(2019)Chen, Sidky, and Ferguson]{Chen2019}
W.~Chen, H.~Sidky and A.~L. Ferguson, \emph{Journal of Chemical Physics}, 2019,
  \textbf{151}, 064123\relax
\mciteBstWouldAddEndPuncttrue
\mciteSetBstMidEndSepPunct{\mcitedefaultmidpunct}
{\mcitedefaultendpunct}{\mcitedefaultseppunct}\relax
\EndOfBibitem
\bibitem[Wu \emph{et~al.}(2018)Wu, Mardt, Pasquali, and Noe]{Wu2018}
H.~Wu, A.~Mardt, L.~Pasquali and F.~Noe, \emph{Advances in Neural Information
  Processing Systems}, 2018, \textbf{31}, 3975--3984\relax
\mciteBstWouldAddEndPuncttrue
\mciteSetBstMidEndSepPunct{\mcitedefaultmidpunct}
{\mcitedefaultendpunct}{\mcitedefaultseppunct}\relax
\EndOfBibitem
\bibitem[Chen \emph{et~al.}(2019)Chen, Sidky, and Ferguson]{Chen2019a}
W.~Chen, H.~Sidky and A.~L. Ferguson, \emph{Journal of Chemical Physics}, 2019,
  \textbf{150}, 214114\relax
\mciteBstWouldAddEndPuncttrue
\mciteSetBstMidEndSepPunct{\mcitedefaultmidpunct}
{\mcitedefaultendpunct}{\mcitedefaultseppunct}\relax
\EndOfBibitem
\bibitem[Bishop(1994)]{Bishop1994}
C.~M. Bishop, \emph{Mixture Density Networks (NCRG/94/004)}, {Aston University}
  technical report, 1994\relax
\mciteBstWouldAddEndPuncttrue
\mciteSetBstMidEndSepPunct{\mcitedefaultmidpunct}
{\mcitedefaultendpunct}{\mcitedefaultseppunct}\relax
\EndOfBibitem
\bibitem[Bishop(2006)]{bishop2006pattern}
C.~M. Bishop, \emph{Pattern Recognition and Machine Learning}, Springer,
  Berlin, 2006\relax
\mciteBstWouldAddEndPuncttrue
\mciteSetBstMidEndSepPunct{\mcitedefaultmidpunct}
{\mcitedefaultendpunct}{\mcitedefaultseppunct}\relax
\EndOfBibitem
\bibitem[Gulrajani \emph{et~al.}(2017)Gulrajani, Ahmed, Arjovsky, Dumoulin, and
  Courville]{Gulrajani2017}
I.~Gulrajani, F.~Ahmed, M.~Arjovsky, V.~Dumoulin and A.~Courville,
  \emph{Advances in Neural Information Processing Systems}, 2017, \textbf{30},
  5768--5778\relax
\mciteBstWouldAddEndPuncttrue
\mciteSetBstMidEndSepPunct{\mcitedefaultmidpunct}
{\mcitedefaultendpunct}{\mcitedefaultseppunct}\relax
\EndOfBibitem
\bibitem[Koltai \emph{et~al.}(2018)Koltai, Wu, No{\'e}, and
  Sch{\"u}tte]{koltai2018optimal}
P.~Koltai, H.~Wu, F.~No{\'e} and C.~Sch{\"u}tte, \emph{Computation}, 2018,
  \textbf{6}, 22\relax
\mciteBstWouldAddEndPuncttrue
\mciteSetBstMidEndSepPunct{\mcitedefaultmidpunct}
{\mcitedefaultendpunct}{\mcitedefaultseppunct}\relax
\EndOfBibitem
\bibitem[Klus \emph{et~al.}(2018)Klus, N{\"u}ske, Koltai, Wu, Kevrekidis,
  Sch{\"u}tte, and No{\'e}]{klus2018data}
S.~Klus, F.~N{\"u}ske, P.~Koltai, H.~Wu, I.~Kevrekidis, C.~Sch{\"u}tte and
  F.~No{\'e}, \emph{Journal of Nonlinear Science}, 2018, \textbf{28},
  985--1010\relax
\mciteBstWouldAddEndPuncttrue
\mciteSetBstMidEndSepPunct{\mcitedefaultmidpunct}
{\mcitedefaultendpunct}{\mcitedefaultseppunct}\relax
\EndOfBibitem
\bibitem[No{\'{e}} and N{\"{u}}ske(2013)]{Noe2013}
F.~No{\'{e}} and F.~N{\"{u}}ske, \emph{Multiscale Modeling and Simulation},
  2013, \textbf{11}, 635--655\relax
\mciteBstWouldAddEndPuncttrue
\mciteSetBstMidEndSepPunct{\mcitedefaultmidpunct}
{\mcitedefaultendpunct}{\mcitedefaultseppunct}\relax
\EndOfBibitem
\bibitem[N{\"{u}}ske \emph{et~al.}(2014)N{\"{u}}ske, Keller,
  P{\'{e}}rez-Hern{\'{a}}ndez, Mey, and No{\'{e}}]{Nuske2014}
F.~N{\"{u}}ske, B.~G. Keller, G.~P{\'{e}}rez-Hern{\'{a}}ndez, A.~S. Mey and
  F.~No{\'{e}}, \emph{Journal of Chemical Theory and Computation}, 2014,
  \textbf{10}, 1739--1752\relax
\mciteBstWouldAddEndPuncttrue
\mciteSetBstMidEndSepPunct{\mcitedefaultmidpunct}
{\mcitedefaultendpunct}{\mcitedefaultseppunct}\relax
\EndOfBibitem
\bibitem[Wu and No{\'e}(2020)]{wu2020variational}
H.~Wu and F.~No{\'e}, \emph{Journal of Nonlinear Science}, 2020, \textbf{30},
  23--66\relax
\mciteBstWouldAddEndPuncttrue
\mciteSetBstMidEndSepPunct{\mcitedefaultmidpunct}
{\mcitedefaultendpunct}{\mcitedefaultseppunct}\relax
\EndOfBibitem
\bibitem[Li \emph{et~al.}(2017)Li, Dietrich, Bollt, and
  Kevrekidis]{li2017extended}
Q.~Li, F.~Dietrich, E.~M. Bollt and I.~G. Kevrekidis, \emph{Chaos: An
  Interdisciplinary Journal of Nonlinear Science}, 2017, \textbf{27},
  103111\relax
\mciteBstWouldAddEndPuncttrue
\mciteSetBstMidEndSepPunct{\mcitedefaultmidpunct}
{\mcitedefaultendpunct}{\mcitedefaultseppunct}\relax
\EndOfBibitem
\bibitem[Andrew \emph{et~al.}(2013)Andrew, Arora, Bilmes, and
  Livescu]{Andrew2013}
G.~Andrew, R.~Arora, J.~Bilmes and K.~Livescu, Proceedings of the 30th
  International Conference on Machine Learning, 2013, pp. 2284--2292\relax
\mciteBstWouldAddEndPuncttrue
\mciteSetBstMidEndSepPunct{\mcitedefaultmidpunct}
{\mcitedefaultendpunct}{\mcitedefaultseppunct}\relax
\EndOfBibitem
\bibitem[Pathak \emph{et~al.}(2018)Pathak, Hunt, Girvan, Lu, and
  Ott]{Pathak2018}
J.~Pathak, B.~Hunt, M.~Girvan, Z.~Lu and E.~Ott, \emph{Physical Review
  Letters}, 2018, \textbf{120}, 24102\relax
\mciteBstWouldAddEndPuncttrue
\mciteSetBstMidEndSepPunct{\mcitedefaultmidpunct}
{\mcitedefaultendpunct}{\mcitedefaultseppunct}\relax
\EndOfBibitem
\bibitem[Goodfellow \emph{et~al.}(2014)Goodfellow, Pouget-Abadie, Mirza, Xu,
  Warde-Farley, Ozair, Courville, and Bengio]{Goodfellow2014}
I.~J. Goodfellow, J.~Pouget-Abadie, M.~Mirza, B.~Xu, D.~Warde-Farley, S.~Ozair,
  A.~Courville and Y.~Bengio, \emph{arXiv preprint arXiv:1406.2661}, 2014,
  https://arxiv.org/abs/1406.2661v1\relax
\mciteBstWouldAddEndPuncttrue
\mciteSetBstMidEndSepPunct{\mcitedefaultmidpunct}
{\mcitedefaultendpunct}{\mcitedefaultseppunct}\relax
\EndOfBibitem
\bibitem[Arjovsky \emph{et~al.}(2017)Arjovsky, Chintala, and
  Bottou]{Arjovsky2017}
M.~Arjovsky, S.~Chintala and L.~Bottou, Proceedings of the 34th International
  Conference on Machine Learning, 2017, pp. 298--321\relax
\mciteBstWouldAddEndPuncttrue
\mciteSetBstMidEndSepPunct{\mcitedefaultmidpunct}
{\mcitedefaultendpunct}{\mcitedefaultseppunct}\relax
\EndOfBibitem
\bibitem[Mirza and Osindero(2014)]{Mirza2014}
M.~Mirza and S.~Osindero, \emph{arXiv preprint arXiv:1411.1784}, 2014,
  https://arxiv.org/abs/1411.1784v1\relax
\mciteBstWouldAddEndPuncttrue
\mciteSetBstMidEndSepPunct{\mcitedefaultmidpunct}
{\mcitedefaultendpunct}{\mcitedefaultseppunct}\relax
\EndOfBibitem
\bibitem[Beauchamp \emph{et~al.}(2011)Beauchamp, Bowman, Lane, Maibaum, Haque,
  and Pande]{beauchamp2011msmbuilder2}
K.~A. Beauchamp, G.~R. Bowman, T.~J. Lane, L.~Maibaum, I.~S. Haque and V.~S.
  Pande, \emph{Journal of Chemical Theory and Computation}, 2011, \textbf{7},
  3412--3419\relax
\mciteBstWouldAddEndPuncttrue
\mciteSetBstMidEndSepPunct{\mcitedefaultmidpunct}
{\mcitedefaultendpunct}{\mcitedefaultseppunct}\relax
\EndOfBibitem
\bibitem[Kingma and Ba(2014)]{kingma2014adam}
D.~P. Kingma and J.~Ba, \emph{arXiv preprint arXiv:1412.6980}, 2014,
  https://arxiv.org/abs/1412.6980v1\relax
\mciteBstWouldAddEndPuncttrue
\mciteSetBstMidEndSepPunct{\mcitedefaultmidpunct}
{\mcitedefaultendpunct}{\mcitedefaultseppunct}\relax
\EndOfBibitem
\bibitem[Goodfellow \emph{et~al.}(2016)Goodfellow, Bengio, and
  Courville]{goodfellow2016deep}
I.~Goodfellow, Y.~Bengio and A.~Courville, \emph{Deep Learning}, MIT Press,
  Cambridge, MA, 2016\relax
\mciteBstWouldAddEndPuncttrue
\mciteSetBstMidEndSepPunct{\mcitedefaultmidpunct}
{\mcitedefaultendpunct}{\mcitedefaultseppunct}\relax
\EndOfBibitem
\bibitem[Lindorff-Larsen \emph{et~al.}(2011)Lindorff-Larsen, Piana, Dror, and
  Shaw]{Lindorff-Larsen2011}
K.~Lindorff-Larsen, S.~Piana, R.~O. Dror and D.~E. Shaw, \emph{Science}, 2011,
  \textbf{334}, 517--520\relax
\mciteBstWouldAddEndPuncttrue
\mciteSetBstMidEndSepPunct{\mcitedefaultmidpunct}
{\mcitedefaultendpunct}{\mcitedefaultseppunct}\relax
\EndOfBibitem
\bibitem[Ramachandran \emph{et~al.}(2017)Ramachandran, Zoph, and
  Le]{ramachandran2017swish}
P.~Ramachandran, B.~Zoph and Q.~V. Le, \emph{arXiv preprint arXiv:1710.05941},
  2017,  https://arxiv.org/abs/1710.05941v1\relax
\mciteBstWouldAddEndPuncttrue
\mciteSetBstMidEndSepPunct{\mcitedefaultmidpunct}
{\mcitedefaultendpunct}{\mcitedefaultseppunct}\relax
\EndOfBibitem
\bibitem[P{\'e}rez-Hern{\'a}ndez \emph{et~al.}(2013)P{\'e}rez-Hern{\'a}ndez,
  Paul, Giorgino, De~Fabritiis, and No{\'e}]{perez2013identification}
G.~P{\'e}rez-Hern{\'a}ndez, F.~Paul, T.~Giorgino, G.~De~Fabritiis and
  F.~No{\'e}, \emph{The Journal of Chemical Physics}, 2013, \textbf{139},
  07B604\_1\relax
\mciteBstWouldAddEndPuncttrue
\mciteSetBstMidEndSepPunct{\mcitedefaultmidpunct}
{\mcitedefaultendpunct}{\mcitedefaultseppunct}\relax
\EndOfBibitem
\bibitem[No{\'e} and Nuske(2013)]{noe2013variational}
F.~No{\'e} and F.~Nuske, \emph{Multiscale Modeling \& Simulation}, 2013,
  \textbf{11}, 635--655\relax
\mciteBstWouldAddEndPuncttrue
\mciteSetBstMidEndSepPunct{\mcitedefaultmidpunct}
{\mcitedefaultendpunct}{\mcitedefaultseppunct}\relax
\EndOfBibitem
\bibitem[N{\"u}ske \emph{et~al.}(2014)N{\"u}ske, Keller,
  P{\'e}rez-Hern{\'a}ndez, Mey, and No{\'e}]{nuske2014variational}
F.~N{\"u}ske, B.~G. Keller, G.~P{\'e}rez-Hern{\'a}ndez, A.~S. J.~S. Mey and
  F.~No{\'e}, \emph{Journal of Chemical Theory and Computation}, 2014,
  \textbf{10}, 1739--1752\relax
\mciteBstWouldAddEndPuncttrue
\mciteSetBstMidEndSepPunct{\mcitedefaultmidpunct}
{\mcitedefaultendpunct}{\mcitedefaultseppunct}\relax
\EndOfBibitem
\bibitem[No{\'e} and Clementi(2015)]{noe2015kinetic}
F.~No{\'e} and C.~Clementi, \emph{Journal of Chemical Theory and Computation},
  2015, \textbf{11}, 5002--5011\relax
\mciteBstWouldAddEndPuncttrue
\mciteSetBstMidEndSepPunct{\mcitedefaultmidpunct}
{\mcitedefaultendpunct}{\mcitedefaultseppunct}\relax
\EndOfBibitem
\bibitem[No{\'e} \emph{et~al.}(2016)No{\'e}, Banisch, and
  Clementi]{noe2016commute}
F.~No{\'e}, R.~Banisch and C.~Clementi, \emph{Journal of Chemical Theory and
  Computation}, 2016, \textbf{12}, 5620--5630\relax
\mciteBstWouldAddEndPuncttrue
\mciteSetBstMidEndSepPunct{\mcitedefaultmidpunct}
{\mcitedefaultendpunct}{\mcitedefaultseppunct}\relax
\EndOfBibitem
\bibitem[P{\'e}rez-Hern{\'a}ndez and No{\'e}(2016)]{perez2016hierarchical}
G.~P{\'e}rez-Hern{\'a}ndez and F.~No{\'e}, \emph{Journal of Chemical Theory and
  Computation}, 2016, \textbf{12}, 6118--6129\relax
\mciteBstWouldAddEndPuncttrue
\mciteSetBstMidEndSepPunct{\mcitedefaultmidpunct}
{\mcitedefaultendpunct}{\mcitedefaultseppunct}\relax
\EndOfBibitem
\bibitem[Schwantes and Pande(2013)]{schwantes2013improvements}
C.~R. Schwantes and V.~S. Pande, \emph{Journal of Chemical Theory and
  Computation}, 2013, \textbf{9}, 2000--2009\relax
\mciteBstWouldAddEndPuncttrue
\mciteSetBstMidEndSepPunct{\mcitedefaultmidpunct}
{\mcitedefaultendpunct}{\mcitedefaultseppunct}\relax
\EndOfBibitem
\bibitem[Chen \emph{et~al.}(2018)Chen, Tan, and Ferguson]{chen2018collective}
W.~Chen, A.~R. Tan and A.~L. Ferguson, \emph{The Journal of Chemical Physics},
  2018, \textbf{149}, 072312\relax
\mciteBstWouldAddEndPuncttrue
\mciteSetBstMidEndSepPunct{\mcitedefaultmidpunct}
{\mcitedefaultendpunct}{\mcitedefaultseppunct}\relax
\EndOfBibitem
\bibitem[Chiavazzo \emph{et~al.}(2017)Chiavazzo, Covino, Coifman, Gear,
  Georgiou, Hummer, and Kevrekidis]{chiavazzo2017intrinsic}
E.~Chiavazzo, R.~Covino, R.~R. Coifman, C.~W. Gear, A.~S. Georgiou, G.~Hummer
  and I.~G. Kevrekidis, \emph{Proceedings of the National Academy of Sciences
  of the United States of America}, 2017, \textbf{114}, E5494--E5503\relax
\mciteBstWouldAddEndPuncttrue
\mciteSetBstMidEndSepPunct{\mcitedefaultmidpunct}
{\mcitedefaultendpunct}{\mcitedefaultseppunct}\relax
\EndOfBibitem
\bibitem[Preto and Clementi(2014)]{preto2014fast}
J.~Preto and C.~Clementi, \emph{Physical Chemistry Chemical Physics}, 2014,
  \textbf{16}, 19181--19191\relax
\mciteBstWouldAddEndPuncttrue
\mciteSetBstMidEndSepPunct{\mcitedefaultmidpunct}
{\mcitedefaultendpunct}{\mcitedefaultseppunct}\relax
\EndOfBibitem
\bibitem[Zheng \emph{et~al.}(2013)Zheng, Rohrdanz, and
  Clementi]{zheng2013rapid}
W.~Zheng, M.~A. Rohrdanz and C.~Clementi, \emph{The Journal of Physical
  Chemistry B}, 2013, \textbf{117}, 12769--12776\relax
\mciteBstWouldAddEndPuncttrue
\mciteSetBstMidEndSepPunct{\mcitedefaultmidpunct}
{\mcitedefaultendpunct}{\mcitedefaultseppunct}\relax
\EndOfBibitem
\bibitem[Krylov \emph{et~al.}(2018)Krylov, Windus, Barnes, Marin-Rimoldi, Nash,
  Pritchard, Smith, Altarawy, Saxe, Clementi, Crawford, Harrison, Jha, Pande,
  and Head-Gordon]{krylov2018perspective}
A.~Krylov, T.~L. Windus, T.~Barnes, E.~Marin-Rimoldi, J.~A. Nash, B.~Pritchard,
  D.~G. Smith, D.~Altarawy, P.~Saxe, C.~Clementi, T.~D. Crawford, R.~J.
  Harrison, S.~Jha, V.~S. Pande and T.~Head-Gordon, \emph{The Journal of
  Chemical Physics}, 2018, \textbf{149}, 180901\relax
\mciteBstWouldAddEndPuncttrue
\mciteSetBstMidEndSepPunct{\mcitedefaultmidpunct}
{\mcitedefaultendpunct}{\mcitedefaultseppunct}\relax
\EndOfBibitem
\bibitem[Wilkins-Diehr and Crawford(2018)]{wilkins2018nsf}
N.~Wilkins-Diehr and T.~D. Crawford, \emph{Computing in Science \&
  Engineering}, 2018, \textbf{20}, 26--38\relax
\mciteBstWouldAddEndPuncttrue
\mciteSetBstMidEndSepPunct{\mcitedefaultmidpunct}
{\mcitedefaultendpunct}{\mcitedefaultseppunct}\relax
\EndOfBibitem
\end{mcitethebibliography}
\bibliographystyle{rsc} 

\end{document}